# Towards a Molecular Level Understanding of the Sulfanilamide-Soil Organic Matter-Interaction

**Ashour A. Ahmed[a,b,c,*], Sören Thiele-Bruhn[d], Peter Leinweber[b,e], Oliver Kühn[a]**

[a] University of Rostock, Institute of Physics, Albert-Einstein-Str. 23-24, D-18059 Rostock, Germany.
[b] Steinbeis GmbH & Co. KG für Technologietransfer, 70174 Stuttgart, Germany.
[c] University of Cairo, Faculty of Science, Department of Chemistry, 12613 Giza, Egypt.
[d] University of Trier, Soil Science, D-54286 Trier, Germany.
[e] University of Rostock, Soil Science, D-18051 Rostock, Germany.

ashour.ahmed@uni-rostock.de   **(Corresponding Author)**
thiele@uni-trier.de
peter.leinweber@uni-rostock.de
oliver.kuehn@uni-rostock.de

## Abstract
In-depth understanding of the sorption mechanisms of organic pollutants, like the antibiotic sulfanilamide (SAA), in soil requires a combined experimental and theoretical approach. Therefore, sorption experiments of SAA on well-characterized samples of soil size-fractions were combined with the modeling of SAA-soil-interaction *via* quantum chemical calculations. Freundlich unit capacities were determined in batch experiments and it was found that they increase with the soil organic matter (SOM) content according to the order fine silt > medium silt > clay > whole soil > coarse silt > sand. The calculated binding energies for mass-spectrometrically quantified sorption sites followed the order ionic species > peptides > carbohydrates > phenols and lignin monomers > lignin dimers > heterocyclic compounds > fatty acids > sterols > aromatic compounds > lipids, alkanes, and alkenes. SAA forms H-bonds through its different polar centers with polar SOM sorption sites. In contrast dispersion and π-π-interactions predominate the interaction of the sulfonamide aromatic ring with the non-polar moieties of SOM. Moreover, the dipole moment, partial atomic charges, and molecular volume of the SOM sorption sites are the main physical properties controlling the SAA-SOM-interaction. The correlation between experimental and theoretical results was established by reasonable estimates of the Freundlich unit capacities from the calculated binding energies. Consequently, we suggest using this approach in forthcoming studies to disclose the interactions of a wide range of organic pollutants with SOM.

# 1. Introduction

Pharmaceutical antibiotics, that are mostly polar and ionizable compounds, have been identified as emerging pollutants. Typically, they reach the soil through contaminated manure from medicated livestock used as fertilizer (Boxall et al., 2004; Thiele-Bruhn, 2003). Sulfonamides, a class of antibiotic substances, are applied in large quantities and were often detected in agricultural soils (Kim et al., 2011). Although knowledge about the extent and kinetics of sulfonamides' sorption in soil accumulated in the past years, uncertainties about sites and mechanisms of sorption still exist (Figueroa-Diva et al., 2010). Consistently, low soil sorption coefficients are determined for sulfonamides (Białk-Bielińska et al., 2012; Figueroa-Diva et al., 2010; Thiele-Bruhn et al., 2004). Sorption substantially increases within hours leading to a decline in extractability (Müller et al., 2012; Wang et al., 2006). The spontaneous and not fully reversible immobilization is partly explained by surface complexation (Figueroa-Diva et al., 2010; Lertpaitoonpan et al., 2009; Schwarz et al., 2012). Furthermore, it is assumed that diffusion and entrapment in micropores of soil sorbents contributes to the strong immobilization (Schwarz et al., 2012; Wang et al., 1993). Sorption of sulfonamides is largely governed by soil organic matter (SOM) while it is subordinate to clay minerals and pedogenic oxides (Figueroa-Diva et al., 2010; Thiele-Bruhn et al., 2004). From experimental findings it was concluded that sorption to SOM is preferred at functional groups of high polarity such as keto, enol, alcoholic and phenolic OH as well as carboxyl groups (Gao and Pedersen, 2005; Thiele-Bruhn et al., 2004). Sorption is assumed to occur *via* ion exchange and ion bridging of charged species as well as van der Waals forces and hydrogen bridges, but possibly also through π-π-interactions of less polar molecular moieties of the neutral species with aromatic ring systems of the sorbent (Gao and Pedersen, 2005; Schwarz et al., 2012; Thiele-Bruhn et al., 2004; Tolls, 2001). It is controversially discussed, though, if hydrophobic partitioning is also relevant (Figueroa-Diva et al., 2010; Lertpaitoonpan et al., 2009). The specific sorption of the sulfonamides is mirrored in non-linear sorption isotherms that are often best described by the Freundlich model (Białk-Bielińska et al., 2012; Lertpaitoonpan et al., 2009; Sanders et al., 2008; Thiele-Bruhn et al., 2004).

Molecular modeling and computational chemistry is a complementary approach, in addition to sorption experiments, to develop a molecular understanding of the binding of pollutants to soil (Gerzabek et al., 2001; Schaumann and Thiele-Bruhn, 2011). Modeling of SOM is not straightforward due to its high variability in the chemical composition, spatial architecture, and multi-phase behavior (Senesi et al., 2009). Most notably, there are different hypotheses concerning the SOM principal structural organization (Schaumann and Thiele-Bruhn, 2011), i.e. macromolecular vs. supramolecular structure (Schaumann, 2006). Several concepts for molecular-scale SOM modeling have been introduced, ranging from (i) complex polymeric models (Schulten et al., 2000; Schulten, 2002) to (ii) the modeling of single functional groups (Aquino et al., 2007, 2009). These models could be criticized because of the huge number of possible combinations for all molecular building blocks into a single macromolecule (i) or due to the narrow selection of functionalities (ii). Therefore, to overcome these problems, recently Ahmed et al. (2014a, 2014b) have developed a new approach for SOM modeling based on SOM characterization by different analytical techniques (Ahmed et al., 2012), which is combined with quantum chemical and molecular dynamics calculations. The model includes a large test set of separate representative systems covering the most relevant functional groups that exist in analytically quantified compound classes (Ahmed et al., 2012). The validity of this model has been proven by experimental adsorption of the non-polar hexachlorobenzene (HCB) on well-characterized soil samples (Ahmed et al., 2014a). The influence of SOM on soil sorption can be ideally determined using (i) soils from one area with similar mineral composition but with different SOM content (Ahangar et al., 2008) and (ii) particle-size fractions of soil, exhibiting different SOM content and composition (Nkedi-Kizza et al., 1983; Schulten et al., 1993).

Complementary to the study on non-polar HCB, the main objective of the present study is exploring the sorption of a polar chemical, i.e. the pharmaceutical antibiotic sulfanilamide (SAA), with SOM at the molecular level. To this end, batch sorption experiments for SAA were performed using particle-size fractions of two soils differing in SOM content and composition due to long-term different fertilization. Furthermore, molecular modeling for the SAA-SOM binding based on the recent SOM model by Ahmed et al. (2014a, 2014b) and quantum mechanical calculations was conducted. Finally, quantitative structure-activity relationship (QSAR) (Nantasenamat et al., 2010) was used to link the SAA-SOM binding to the physicochemical properties of the SOM functional groups.

## 2. Material and methods

### 1.1. Soil samples and particle-size fractionation

In order to test soil samples that differ specifically in SOM, topsoil samples were taken from the Ap horizon (0 to 20 cm depth) of a haplic Phaeozem from the long-term 'Eternal Rye Cultivation experiment' at Halle (Saale), Germany (Kühn, 1901; Schmidt et al., 2000). Representative samples from two differently fertilized plots, i.e. the unfertilized treatment (U), and a plot that received farmyard manure from 1878 until sampling date in 2000 (FYM) were used (Schmidt et al., 2000). Both soil samples had a similar mineral composition with illite, smectite and mixed layer minerals predominating in the clay fraction (<2 µm) but differed substantially in SOM content (Leinweber and Reuter, 1989). The samples were air dried and sieved (<2 mm) prior to experiments. Additionally, both samples U and FYM were separated each into five particle-size fractions (for details, see the supplementary information (SI)), i.e. sand (2000–63 µm), coarse silt (63–20 µm), medium silt (20–6.3 µm), fine silt (6.3–2 µm), and clay (<2 µm) (Leinweber et al., 2009; Amelung et al., 1998, Schmidt et al., 1999). Selected general characteristics of the whole topsoil samples and their respective particle-size fractions such as organic carbon (OC), nitrogen (N), sulfur (S), cation exchange capacity (CEC), pH, and content of the pedogenic oxides (iron, aluminum, and manganese) extracted by a dithionite-citrate-bicarbonate solution ($Fe_{dith}$, $Al_{dith}$, and $Mn_{dith}$) are listed in Table S1 in SI.

### 1.2. Pyrolysis field ionization mass spectrometry (Py-FIMS)

For each topsoil and particle-size fraction sample, the Py-FI mass spectrum, containing the marker signals of important SOM chemical compounds in the mass range of 55 to 500 au, was obtained (more details are given in SI). According to well established modes of spectra interpretation (Ahmed et al., 2012; Schulten and Leinweber, 1999) the particular ion intensity (I) of each (1) carbohydrates with pentose and hexose subunits (CHYDR), (2) phenols and lignin monomers (PHLM), (3) lignin dimers (LDIM), (4) lipids, alkanes, alkenes, bound fatty acids, and alkyl monoesters (LIPID), (5) alkyl aromatics (ALKY), (6) non-peptidic (e.g., nitriles, N-heterocyclic compounds) N-containing compounds (referred to as N-containing compounds) (NCOMP), (7) sterols (STEROL), (8) peptides (PEPTI), and (9) free fatty acids (FATTY) was calculated. Furthermore, total ion intensity ($I^{tot}$) that is the sum of ion intensities of all recorded marker signals was calculated for each sample.

### 1.3. Adsorption experiment

Adsorption of SAA (purity ≥ 99.0%, Sigma, Taufkirchen, Germany; for details see the 2D structure of SAA in Figure S1 in SI) was determined in batch trials according to OECD guideline 106 (OECD, 2000) and based on previous studies (Ahmed et al., 2015; Thiele-Bruhn et al., 2004). All samples were done in triplicate. For each replicate, 5.0 g of air-dried soil or soil fraction was weighed into 75-mL glass centrifuge tubes and spiked with SAA in five concentrations (0, 0.58, 5.81, 58.1, and 232.3 µmol/kg). To this end, SAA was dissolved in <0.5 mL methanol. Methanol was found not to affect sorption experiments up to at least 0.5 vol%. After the solvent was allowed to evaporate for 1 h, 0.01 $M$ $CaCl_2$ was added in a soil-to-solution ratio of 1:2.5 (w/v). Samples were shaken on an end-over-end rotary shaker at 15 rpm for 16 h at 22°C in the dark prior to centrifugation for 30 min at 1700 ×$g$. The shaking time of 16 h is ample to sufficiently reach sorption equilibrium (Thiele-Bruhn et al., 2004). Previous tests showed that > 90% of the added sulfonamide can be recovered using harsh extraction methods (Thiele-Bruhn and Aust, 2004) and that biodegradation in soil is negligible even on a long-term (Rosendahl et al., 2011).

The supernatants from the sorption experiments were directly analyzed for SAA using HPLC. A Hewlett-Packard (Palo Alto, CA) 1050 HPLC system equipped with a wavelength programmable UV detector (HP 1050) and a fluorescence detector (HP 1046A) was used. A 250×4.6 mm Nucleosil 100-5-C18 reversed-phase column served as stationary phase (Macherey-Nagel, Düren, Germany). The mobile phase consisting of (A) 0.01 $M$ $H_3PO_4$ and (B) methanol was delivered in a gradient program at a flow rate of 1.0 mL/min. Using injection volumes of 10 µL, SAA was determined with UV detection at 265 nm and fluorescence detection at 276/340 nm. The detection limit for SAA was 0.03 µmol/L (Thiele-Bruhn and Aust, 2004).

The non-linear Freundlich isotherm (Eq. 1) were fitted to the data of the total SAA sorbate concentration associated with the sorbent ($q$, µmol/kg) and the total SAA concentration remaining in the equilibrium solution ($c_w$, µmol/L) using the CFIT software for non-linear regression (Helfrich, 1996):

$$q = K_f \times c_w^n \qquad (1)$$

with $K_f$ the Freundlich unit capacity coefficient (µmol$^{1-n}$ L$^{1/n}$/kg) and $n$ the dimensionless Freundlich exponent indicating sorption non-linearity.

### 1.4. SOM modeling and quantum chemical calculations

We applied a previously introduced SOM model (Ahmed et al., 2014a) that is based on detailed molecular analyses by Py-FIMS and XANES (Ahmed et al., 2012). Thereby, SOM was modeled by a set of representative systems covering a broad range of functional groups as well as compound classes of SOM. Our model, see Figure S2 in SI, included PHLM represented by phenol, catechol, and 3,4,5-trimethoxy cinnamic acid (lignin monomer); ALKY represented by benzene, methylbenzene, ethylbenzene, naphthalene, and ethylnaphthalene; CHYDR represented by glucose in the open and cyclic forms; PEPTI represented by glycine and penta-glycine; NCOMP represented by ethylnitrile, pyrrole and pyridine; and LIPID represented by short- and long-chain alkane and conjugated alkene. Furthermore, effect of the free fatty acids (FATTY) on binding of SAA to soil was compiled from the modeled carboxylic acid and long-chain alkane and alkene functional groups. Moreover, binding to sterols (STEROL) was investigated by including the hydroxyl group in methanol combined with the long-chain alkane and alkene. The impact of the lignin dimers (LDIM) was assembled from the modeled lignin monomer. To study the effect of the SOM polarity, the same model included different carbonyl functional groups such as acetamide, acetaldehyde, dimethylketone, and methylacetate; amine like methylamine, and aniline; and quinone. To study the effect of ions, protonated methylamine as a positively charged system and acetate anion as a negatively charged system were included in the model.

In soil, there are multiple interactions between the soil components e.g. SOM-SOM, SOM-soil minerals, SOM-xenobiotics, soil minerals-xenobiotics interactions, and so on. Since we are focusing on sorption of SAA to SOM, we considered here the interaction or binding of SAA to the molded SOM fragments. Therefore, 1:1 complex formation between SAA and each individual modeled SOM fragment was assumed. Other complexes such as 1:2 or 2:1 of SAA-SOM-complexes were not considered in the current contribution. For each 1:1 complex, the initial geometries were constructed by selecting the expected preferential binding situations between SAA and the representative SOM fragment. The different initial geometries for each complex were fully geometry optimized in gas phase. In case this resulted in more than one configuration the most stable one has been selected. To calculate the binding energy ($E_{B_i}$) of SAA to the SOM fragments, full geometry optimization was performed for each SAA-SOM-complex as well for all individual species (SAA and each SOM system) in the gas phase. Since the aqueous soil solution is an important factor controlling the SAA-SOM-interaction, it was simulated by a continuum solvation approach. Solvation by water has been incorporated *via* implicit treatment through the conductor-like screening model (COSMO) (Schäfer et al., 2000). Analogous to the gas phase, full geometry optimization was performed for all species (SAA, each SOM system, and each SAA-SOM-complex) using COSMO. All calculations have been performed using the TURBOMOLE program package (Turbomole v6.4, 2012).

The binding energies of SAA to the SOM systems in these complexes were calculated as the difference between the total energies of the complex and the individual molecules.

$$E_{B_i} = E_{(SAA-i)complex} - (E_{SAA} + E_i) \qquad (2)$$

where, $E_{B_i}$ is the binding energy of SAA to the SOM system i, $E_{(SAA-i)complex}$ is the energy of the complex of SAA with the system i, $E_{SAA}$ is energy of SAA, and $E_i$ is energy of the system i.

The interaction of SAA with the SOM representative systems has been studied by density functional theory (DFT) calculations. Here, the Becke, three-parameter, Lee-Yang-Parr hybrid functional (B3LYP) (Becke, 1988; Lee et al., 1988) has been used together with a 6-311++G(d,p) basis set. Dispersion corrections are accounted for by employing the empirical D3 approach by Grimme and coworkers (Grimme et al., 2011). The effect of the basis set superposition error (BSSE) has been corrected using the standard protocol (Jansen and Ros, 1969).

## Quantitative activity-structure relationship (QSAR)

QSAR analysis has been performed to correlate the calculated binding energy ($E_B$), of SAA to SOM representative systems, with the relevant calculated physical parameters of the SOM representative systems. Various physical parameters (descriptors) that were expected to have an influential role in the binding process were selected. Among the selected descriptors, the following ones characterizing the test systems showed valuable contribution to $E_B$: The dipole moment ($P_1$), quadrupole moment ($P_2$), anisotropy ($P_3$), sum of the partial charges on O atoms ($P_4$), sum of partial charges on C+O+N atoms ($P_5$), molecular-mass ($P_6$), sum of the partial charges on C atoms ($P_7$), and molar volume ($P_8$). These physical properties were correlated to the binding energies *via* the following equation.

$$E_B = C_0 + \sum_{i=1}^{8} C_i * P_i \qquad (3)$$

The coefficients $C_0$ to $C_8$ were determined using multiple-linear regression. In addition, selected statistical parameters were calculated such as sum of squares due to the error (SSE), sum of squares due to the regression (SSR), sum of total squares (SST), mean of squares due to the error (MSE), mean of squares due to the regression (MSR), and mean of total squares (MST). Also, $R^2$ (which is equal to SSR/SST) and adjusted $R^2$ (which is equal to 1-MSE/MST) were calculated which are proportional to the total variation. Finally, $F_{statistics}$ (which is equal to MSR/MSE) that measures significance of the model describing the data was calculated.

## 3. Results and discussion
### 1.5. Sorbent properties and SAA adsorption

The two whole soil samples and the corresponding particle-size fractions showed clear and significant ($p<0.05$) differences in SOM related properties, i.e. OC and N content (Table S1 in SI). The pedogenic oxides content and CEC largely increased with decreasing particle-size. In the Py-FI mass spectra, $I^{tot}$ of the bulk soil samples and soil fractions reflected the differences in OC content (see Table S2 and compare with Table S1 in SI). These results confirm similar data of a previous study on particle-size fractions from samples taken in 1986 at the Eternal Rye Cultivation experiment (Schulten and Leinweber, 1991). Also the differences in organic matter content and CEC were clearly related to the long-term addition of organic fertilizer to the plot FYM. Differences in pedogenic oxides (Table S1 in SI) and SOM composition (Table S2 in SI) among particle-size fractions confirm data compiled in the review by Schulten and Leinweber (2000). However, $I^{tot}$ was unusually large in the fertilized whole soil sample (22.393 x $10^6$ counts/mg). This can be explained by sample heterogeneity that may have resulted in undesired enrichment of easily pyrolyzed manure remnants in the small subsample taken for Py-FIMS as is indicated by the large ion intensity from sterols. Thus, this sample was considered as outlier in the whole sample set.

In the particle-size fractions, unbiased $I^{tot}$ decreased in the order fine silt > medium silt > clay > coarse silt and sand (Table S2 in SI). The order of ion intensities (I) varied among the different compound classes. Carbohydrates had their largest proportions in the sand and clay fractions. Decreasing proportions with decreasing particle-size were observed for lignin dimers, lipids and alkylaromatics whereas the proportions of N-compounds and peptides showed the opposite trend. For the proportions of other compound classes no such clear trends occurred with particle-size. Fertilization with farmyard manure increased the proportions of lignin dimers and sterols in clay, coarse silt and bulk soil. These different trends document the different SOM composition in the investigated soil samples and fractions, although total differences in the percentages of the various compound classes were small (Table S2 in SI).

Correspondingly, sorption of SAA differed among whole soil samples and particle-size fractions. Sorption was low with the Freundlich unit capacity coefficient ($K_f$) ranging from 0.16 to 13.65 and mostly non-linear with the exponent $n$ ranging from 0.5 to 1 (Table 1). Overall, coefficients of non-linear curve fit of SAA sorption to the different samples declined in the sequence fine silt > medium silt > clay > whole soil > coarse silt > sand. Sorption of SAA ($K_f$) to soil samples increased with increasing SOM content. Significant correlations were established for $K_f$ with the elemental indicators of SOM, i.e. OC (r=0.84), N (r=0.61), and S (r=0.73) content, confirming the well-known relevance of SOM for sulfonamide sorption. With respect to the soil minerals, low correlation coefficients were obtained for $Al_{dith}$ (r=0.08), $Fe_{dith}$ (r=0.13), and $Mn_{dith}$ (r=0.16) with $K_F$. This indicates a subordinate contribution of soil mineral colloids to the adsorption of SAA compared to that of SOM (Figueroa-Diva et al., 2010; Lertpaitoonpan et al., 2009; Thiele-Bruhn et al., 2004). Even more, SAA sorption to the clay size fractions was much smaller than expected from the content of organic and mineral sorbents.

In total, sorption of SAA to soil was rather low, which confirms previous findings (Figueroa-Diva et al., 2010; Thiele-Bruhn et al., 2004). The in part strong sorption non-linearity is interpreted as an indicator of site-specific sorption resulting in non-ideal sorption behavior (Pignatello et al., 2006a). Because the pH of the soil samples and fractions was in the range of 5.2 to 6.2 (Table S1), only the neutral SAA molecule occurred (neutral species fraction ≥ 99.99 %); thus, sorption of SAA was not affected by pH. Retention of SAA in soil seemingly did not depend on the content of mineral sorbents but was governed by the total amount of SOM (Figueroa-Diva et al., 2010; Lertpaitoonpan et al., 2009). Furthermore, the smaller sorption of SAA to the clay-size fractions compared to fine silt and medium silt fractions revealed that sorption further varied with the different molecular composition of SOM in the different soils samples and size fractions, as will be discussed in the following, and with the number and availability of sorption sites. It has been previously shown that the association of humic substances with surfaces of other sorbents such as clay leads to a reduction of the specific surface area and sorptive properties through blocking of micro- and nanopores (Kaiser and Guggenberger, 2003; Pignatello et al., 2006b). Especially the reduced accessibility of pores might be relevant for the sorption of SAA; a previous study showed the relevance of such cavities for SAA sorption (Ahmed et al., 2015).

Sorption experiments showed that SAA sorption to soil depends on but is not fully explained by SOM quantity. Obviously, SAA adsorption also depends on SOM quality and its inhomogeneity among particle-size fractions. This was further evidenced by normalizing the Freundlich unit capacity to the OC content of the respective soil samples and fractions (see Table 1 and Table S3 in SI). Even the normalized sorption coefficients differed largely by a factor of up to 5.6 indicating that SAA sorption is not only governed by the SOM content but also SOM composition.

Table 1. Parameters of the Freundlich isotherm fitting to the adsorption data of SAA to two differently fertilized soils and their particle-size fractions.

| soil | fraction | $K_f$ $\mu mol^{1-1/n} L^{1/n}/kg$ | $K_{OC}$ | $n$ | SD | $R^2$ |
|---|---|---|---|---|---|---|
| unfertilized soil (U) | whole soil | 0.97 | 96.04 | 0.81 | 5.21 | 0.88 |
| | sand | 0.16 | 61.54 | 0.89 | 2.57 | 0.65 |
| | coarse silt | 0.33 | 48.53 | 1.02 | 0.45 | 1.00 |
| | medium silt | 9.92 | 271.78 | 0.62 | 2.30 | 1.00 |
| | fine silt | 12.37 | 241.60 | 0.66 | 1.38 | 1.00 |
| | clay | 3.10 | 67.25 | 0.77 | 5.38 | 0.97 |
| fertilized soil (FYM) | whole soil | 1.15 | 79.31 | 0.82 | 5.50 | 0.91 |
| | sand | 0.91 | 182.00 | 0.53 | 1.33 | 0.91 |
| | coarse silt | 0.48 | 82.76 | 0.88 | 4.22 | 0.84 |
| | medium silt | 8.82 | 182.23 | 0.67 | 4.59 | 0.99 |
| | fine silt | 13.65 | 176.36 | 0.85 | 0.58 | 1.00 |
| | clay | 5.71 | 84.72 | 0.72 | 7.18 | 0.97 |

## 1.6. Quantum chemical modeling

The gas phase equilibrium geometries of the complexes of SAA with the representative SOM systems are shown in Figure 1. The complexes are numbered according to increasing gas phase binding energy of SAA with the SOM systems. All the calculated binding energies were corrected by removing BSSE using the counterpoise correction (for details, see Figure S3 in SI). In general, SAA follows different modes of interaction with the SOM representative systems due to its various active centers. Specifically, it comprises five negative (2 N atoms, 2 O atoms, and an aromatic ring) in addition to four positive centers of interaction (4 H atoms of amine groups). For this reason, SAA has the ability to interact with the polar compounds through its partially charged N, O, and H atoms giving rise to H-bond formation. Further, it can interact through its electron density in the aromatic ring with the non-polar compounds. There are no covalent bonds observed between SAA and the SOM model set. For most complexes, SAA forms either one or two H-bond(s) with the polar SOM systems. The length of the observed H-bonds for SAA-SOM-interaction varies from 1.85 to 2.85 Å. Dispersion interaction is observed to be predominant for those systems that have no polar functional group. More details about the role of dispersion interaction in formation and stability of the SAA-SOM-complexes can be found in Figure S4 in SI.

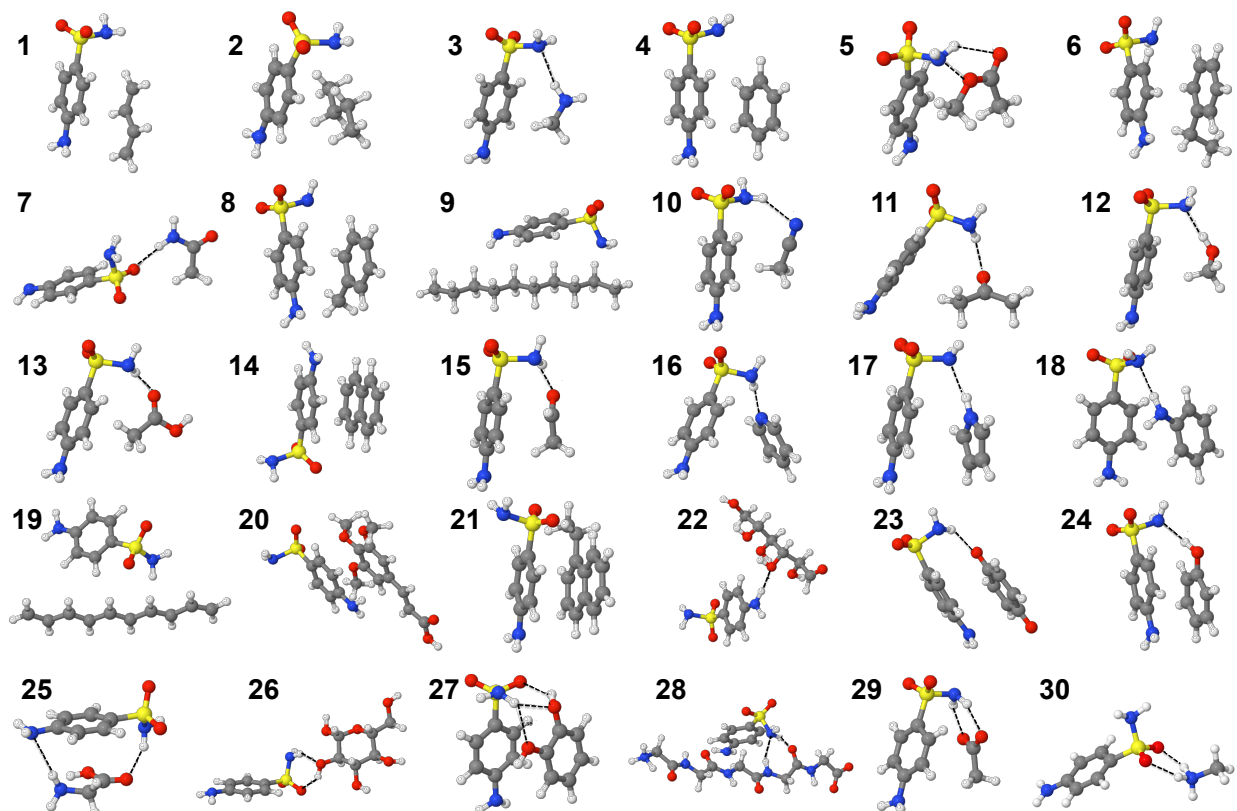

Figure 1. Optimized geometries of SAA-SOM-complexes in gas phase at DFT/D3/B3LYP/6-311++G(d,p) level of theory.

Let us focus onto the effect of SOM quality (i.e. its chemical composition) to address the question "How SOM composition controls SAA-SOM-interaction?". In general, Figure 2 indicates that the polarity of the SOM systems is an important factor leading to increase of the binding energy of SAA to SOM. Apparently, SAA binds to the hydrophilic as well as the charged compounds stronger than to the hydrophobic ones. A detailed inspection shows that binding of SAA to the modeled SOM systems in Figure 2 can be subdivided into five sets. Set **I**, which comprises the modeled compounds from **1** to **9**, represents the lowest binding energies with SAA. This set involves non-polar aliphatic compounds such as short-chain alkene (**1**, -3.9 kcal/mol) and alkane (**2**, -4.4 kcal/mol), and long-chain alkane (**9**, -6.8 kcal/mol) as well as non-polar aromatic compounds such as benzene (**4**, -5.1 kcal/mol), ethylbenzene (**6**, -6.1 kcal/mol) and methylbenzene (**8**, -6.7 kcal/mol). This set mainly comprises hydrophobic compounds except three hydrophilic systems that are amine (**3**, -4.8 kcal/mol), ester (**5**, -5.4 kcal/mol), and amide (**7**, -6.6 kcal/mol). Set **II** (**10**-**15**) represents the hydrophilic systems containing one functional group that are nitrile (**10**, -7.1 kcal/mol), ketone (**11**, -7.3 kcal/mol), alcohol (**12**, -7.6 kcal/mol), carboxylic acid (**13**, -7.7 kcal/mol), aldehyde (**15**, -8.3 kcal/mol). Further, there is one hydrophobic system that is naphthalene (**14**, -8.2 kcal/mol). Set **III** (**16**-**18**) represents pyridine (**16**, -8.4 kcal/mol), pyrrole (**17**, -8.6 kcal/mol), and aniline (**18**, -8.7 kcal/mol). This set can be classified as group for N-heterocyclic and aniline compounds. Set **IV** (**19**-**27**) represents those hydrophilic systems that contain many polar functional groups in addition to the hydrophobic compounds with high electron density. The hydrophobic compounds in this group are the long-chain conjugated alkene (**19**, -8.9 kcal/mol) and ethylnaphthalene (**21**, -9.5 kcal/mol). The lignin monomer (3,4,5-trimethoxy cinnamic acid, **20**, -8.9 kcal/mol), glucose in open (**22**, -9.8 kcal/mol) and cyclic (**26**, -12.8 kcal/mol) forms, quinone (**23**, -10.8 kcal/mol), phenol (**24**, -11.0 kcal/mol), glycine (**25**, -12.2 kcal/mol), and catechol (**27**, -14.7 kcal/mol) are the hydrophilic molecular systems included in this set. Set **V** (**28**-**30**) comprises the representative SOM molecular systems of the highest binding energies with SAA. It is related to peptides that are represented by penta-glycine (**28**, -22.1 kcal/mol), and the charged systems such as acetate anion (**29**, -32.7 kcal/mol) and charged amine (**30**, -40.1 kcal/mol).

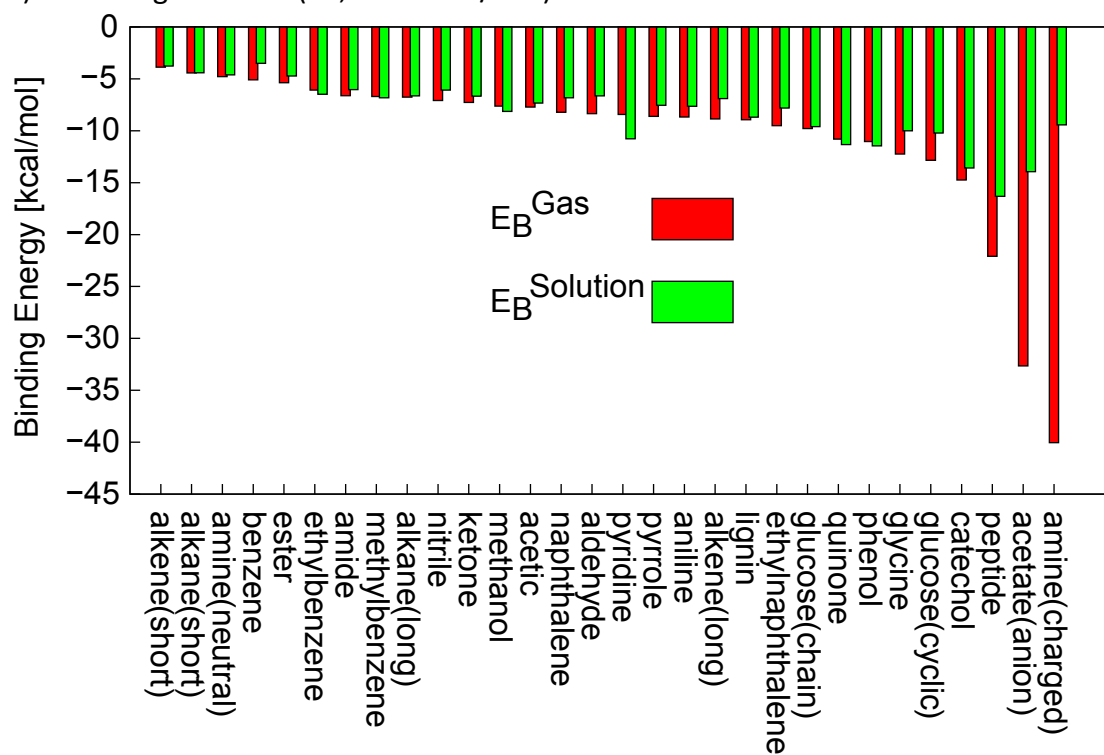

Figure 2. Binding energies for SAA with the representative SOM systems, in the SAA-SOM-complexes shown in Figure 1, calculated at the B3LYP/D3/6-311++G(d,p) level of theory in gas phase (red) as well as in solution (green).

A close examination of Figures 1 and 2 reveals that SAA binds to the aliphatic functional groups in the order long-chain conjugated alkene (**19**) > aldehyde (**15**) > carboxylic acid (**13**) > alcohol (**12**) > ketone (**11**) > nitrile (**10**) > long-chain alkane (**9**) > amide (**7**) > ester (**5**) > amine (**3**) > short-chain alkane (**2**) > short-chain alkene (**1**). This indicates that SAA binds to the hydrophobic systems of longer chain stronger than to those with shorter chain. Moreover, as the electron density increases on these hydrophobic systems the binding energy to SAA increases too. For polar systems, it could be expected that functional groups containing O atom(s) bind stronger to SAA than those containing N atom(s), for example, alcohol (**12**) and amine (**3**). For aromatic compounds, SAA binds in the order catechol (**27**) > phenol (**24**) > ethylnaphthalene (**21**) > 3,4,5-trimethoxy cinnamic acid (**20**) > aniline (**18**) > pyrrole (**17**) > pyridine (**16**) > naphthalene (**14**) > methylbenzene (**8**) > ethylbenzene (**6**) > benzene (**4**). This indicates that SAA binds to N-heterocyclic compounds (**16** and **17**) stronger than to alkylated benzene (**4**, **6**, and **8**). Moreover, SAA binds to the polycyclic aromatic rings (like the substituted (**21**) and non-substituted (**14**) naphthalene) stronger than monocyclic aromatic rings (like the substituted (**6**, **8**) and non-substituted (**4**) benzene). SAA binds to naphthalenes stronger than to benzenes, but the interaction with SAA exceeds that of naphthalenes if benzene is substituted by strong electron donating functional groups (such as OH, OCH$_3$, and NH$_2$). Similarly to aliphatic compounds, SAA binds to aromatic compounds with functional groups containing O atom(s) stronger than to those containing N atom(s), for example, phenol (**24**) and aniline (**18**). Further, SAA binds to peptides (penta-glycine, **28**) stronger than to carbohydrates (glucose monomer, **26**).

The binding energies for the individual SOM systems with SAA were combined into average binding energies for the corresponding compound classes (see Table S4 in SI). The average binding energies (in kcal/mol) showed that SAA binds to SOM constituents in the order cationic species (-40.1) > anionic species (-32.7) > PEPTI (-22.1) > CHYDR (-12.8) > PHLM (-11.6) > LDIM (-8.9) > NCOMP (-8.1) > FATTY (-7.7) > STEROL (-7.6) > ALKY (-7.1) > LIPID (-6.0). This compilation confirms that SAA binds to the charged compounds as well as the hydrophilic molecular systems stronger than to the hydrophobic ones.

Coming to the environmentally more important effect of soil solution on SAA-SOM-interaction, the COSMO calculations showed that the solvation does not affect significantly the geometry of SAA-SOM-complexes (see Figure S5 in SI). Solvation process decreased the binding energies between SAA and the SOM systems for most of the SAA-SOM-complexes compared to gas phase cases (Figure 2). This is due to stabilization of the individual components (SAA and the SOM systems) by water in addition to destabilization of the SAA-SOM-complexes by water. The main reason for this destabilization is that the sum of the solvent accessible area for the individual components (SAA and the SOM systems) is larger than that for the SAA-SOM-complexes. Especially this effect is strong in case of charged or highly hydrophilic SOM system (see the strong decrease of the binding energy for the charged amine (-9.4 kcal/mol) and acetate (-13.9 kcal/mol), and also for peptide (-16.3 kcal/mol) in Figure 2). Nevertheless, the overall picture in presence of water showed an analogous trend in binding of SAA to SOM compared to the gas phase case. Presently, SAA binds to the charged and extremely polar molecular systems stronger than the hydrophobic and lowly polar ones. Regarding the SOM compound classes, SAA binds to PEPTI (-16.3 kcal/mol) > PHLM (-11.2) > CHYDR (-10.2) > LDIM (-8.7) NCOMP = STROL (-8.1) > FATTY (-7.3) > ALKY (-6.3) > LIPID (-5.4). More details about the effect of soil solution on the SAA-SOM-interaction can be shown in SI.

### 1.7. Comparison between experiment and theory

The first goal in this section is how to move from the simple case to the complex one i.e. from the calculated binding energy of SAA to the SOM fragment to an approximated binding energy of SAA to the whole SOM. Here we would simply mention that collection of special SOM fragments would give rise to a certain SOM building block. Similarly, collections of the different SOM building blocks will build the whole SOM. Based on the previous sentences, one can calculate the binding energy of SAA to certain SOM building block as the average of the different calculated binding energy values of SAA to the SOM fragments in this SOM building block (see Eq. 4, these binding energies were collected in Table S4 in SI).

$$E_B^j = \frac{\sum_i E_{B\,i}}{n} \qquad (4)$$

where $E_B^j$ is the binding energy of SAA to certain SOM building block j, $E_{B\,i}$ is the binding energy of SAA to certain SOM fragment i, and $n$ is number of the modeled SOM fragments in the SOM building block j.

Having a well-characterized soil sample containing particular proportions of SOM building blocks, one can estimate that the binding strength of SAA to its whole SOM depends on the proportion of each SOM building block and also on the binding strength of SAA to each SOM building block. Therefore, one can formulate the binding energy of SAA to whole SOM (i.e. the total binding energy) as the sum of the binding energies of SAA to the SOM building blocks weighted by their ion intensities that determined by Py-FIMS (see Table S2 in SI). Then it is allowed now to write the following equation:

$$E_B^{tot} = \sum_j E_B^j * I_j \qquad (5)$$

where $E_B^{tot}$ is the binding energy of SAA to the whole SOM, $I_j$ is the Py-FIMS ion intensity of particular SOM building blocks j.

Since we have now something theoretically ($E_B^{tot}$) that can express the binding strength of SAA to the whole SOM, one may examine its link to the investigated experimental binding strength ($k_f$). Based on Eq. 5, $E_B^{tot}$ was calculated for every soil sample used in the current contribution. Therefore, we correlated the calculated binding energy of SAA to the whole SOM ($E_B^{tot}$) to the corresponding Freundlich unit capacity ($k_f$) that is the second goal for this section. Here the best fitting was obtained via linear relationship between $E_B^{tot}$ and $k_f$, i.e.

$k_f = A * E_B^{tot} + B$ (6)

Using the gas phase binding energies one obtains

$k_f = -0.062\, E_B^{tot} - 0.411$ (7)

with high correlation coefficient (r=0.94) and $R^2$=0.88 (see Figure 3). Further, the calculated binding energies in solution yield the following equation (r=0.95 and $R^2$=0.89)

$k_f = -0.072\, E_B^{tot} - 0.425$ (8)

Both equations already show a reasonable agreement between experimental and theoretical data. This agreement validates various assumptions in our experimental-theoretical-approach: the representativeness of the Py-FI mass spectra for the SOM composition in terms of quantity and quality (Ahmed et al., 2012), the assignment of the *m/z* recorded to compound classes (Leinweber et al., 2009), the representativeness and validity of the chosen molecular subunits in the SOM model (Ahmed et al., 2014a), and the validity of the chosen computational chemistry method to describe the pollutant-SOM-interaction(Ahmed et al., 2014a, 2014b). Thus, the Freundlich unit capacity can be estimated from the calculated binding energy.

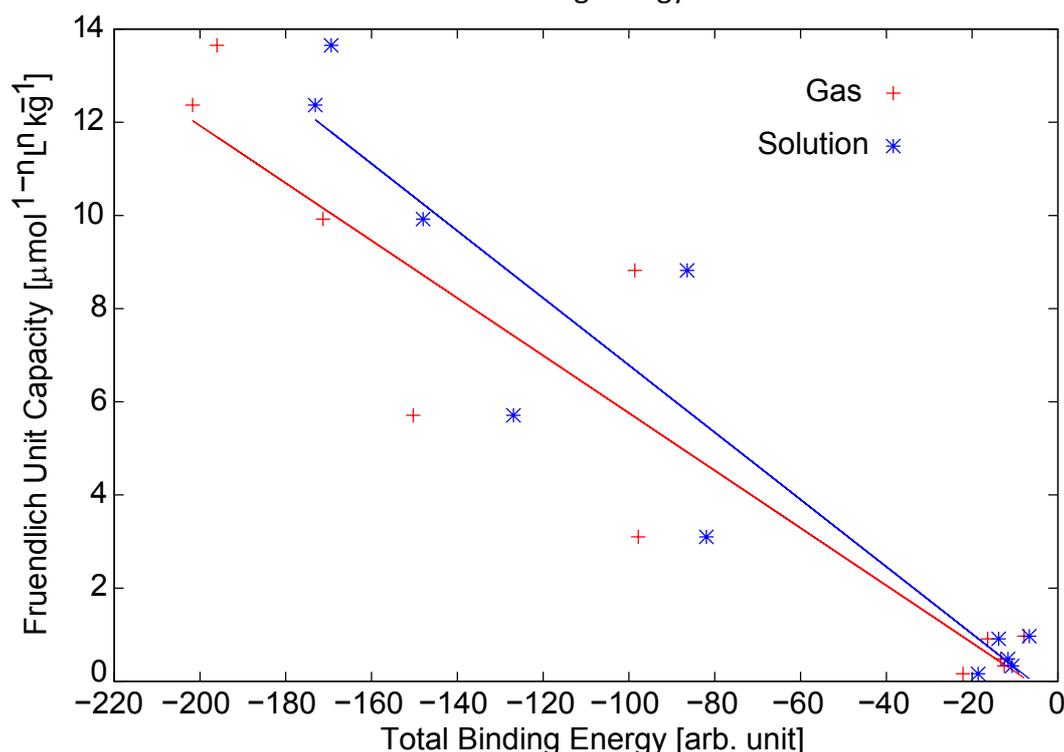

Figure 3. Correlation of the Freundlich unit capacities with the approximated total binding energies of SAA with the soil samples and their fractions in gas phase (red) and in solution (blue) excluding the $k_f$ value for the whole fertilized soil sample.

### 1.8. Comparison with a non-polar pollutant

Given the success of our molecular-computational model for describing SAA-SOM-interactions, it is instructive to compare our current results with that published for binding of hexachlorobenzene (HCB) to the same SOM model (Ahmed et al., 2014a). In gas phase, Figure S6 in SI showed that 15 systems, related to hydrophobic aliphatic and aromatic compounds, bind to HCB stronger than to SAA. For the charged systems and the molecular systems of high polar character, SAA binds stronger than HCB. For those systems having stronger binding to HCB, the differences between the binding energies for HCB and SAA complexes exceeded that for the systems having stronger binding to SAA. This means that in general HCB binds to soil stronger than SAA. Also this can be reflected from the averaged binding energies for the SOM compound classes. HCB binds stronger than SAA to PHLM, LDIM, ALKY, LIPID, FATTY, STROL, and NCOMP. In contrast, SAA binds stronger than HCB to PEPTI and CHYDR.

Including COSMO as a solvation medium model, HCB binds to 13 systems stronger than SAA (see Figure S7 in SI). Compiling these data to averaged binding energies yield stronger binding for HCB to LDIM, ALKY, LIPID, FATTY, and STEROL than for SAA. In contrast, SAA binds stronger than HCB to the hydrophilic compound classes including PEPTI, CHYDR, and NCOMP. Similar binding energies are observed for interaction of HCB and SAA with PHLM compound class that have both the hydrophobic and hydrophilic characters. Finally, these details can be summarized in general into stronger binding for HCB to soil or its surface than that for SAA. This typically agrees with the experimental findings that explored that sorption of the hydrophobic HCB to soil is stronger than sorption of the hydrophilic SAA (Ahmed et al., 2015).

### 1.9. Quantitative activity-structure relationship (QSAR)

For gas phase, the coefficients of Eq. 3 were determined and given in the following equation.

$$E_B(\text{Gas}) = -6.140 - 0.034\, P_1 - 0.121\, P_2 - 0.286\, P_3 + 0.058\, P_4 - 0.457\, P_5 - 0.062\, P_6 - 0.422\, P_7 + 0.027\, P_8 \qquad (9)$$

The estimated binding energies, $E_B(\text{Gas})$, in Eq. 9 were plotted versus the calculated ones (see Figure 4). The fitted parameters of this equation proved efficiency of this generated equation (for details see Table S5 in SI). Hence, the most correlated and contributed descriptors to the binding energy are the anisotropy ($P_3$), dipole moment ($P_1$), sum of the partial charges on C+O+N atoms ($P_5$), and quadrupole moment ($P_2$). This provides evidence for the dependence of the SAA-SOM-interaction on the polarity, charge, and orientation of the interacting SOM fragment to SAA. Moreover, larger absolute values of the dipole moment and/or the anisotropy and/or sum of the partial charges on C+O+N atoms and/or quadrupole moment of SOM consequently result in stronger binding of SAA to SOM systems.

The dependence on the molar volume of SOM fragments points to the importance of the SOM subjected surface area in this interaction. This in turn points to the role of dispersion in this type of interaction. Therefore, the outcome of the QSAR gives evidence for the dual nature of SAA in its interaction with SOM. It strongly interacts with the polar compounds but also does moderately interact with the non-polar ones. This agrees with experimental findings, that SAA and other sulfonamides were sorbed to oligomerized vanillin; sorption was preferred to the O-containing moieties in addition to π-π-interactions with the aromatic ring (Schwarz et al., 2012).

By including effect of water as a solvent surrounding SAA and SOM in their complexes, the following equation was obtained.

$$E_B(\text{COSMO}) = -6.121 + 0.008\, P_1 - 0.148\, P_2 - 0.300\, P_3 + 0.602\, P_4 - 1.300\, P_5 - 0.046\, P_6 \qquad (10)$$

Efficiency of Eq. 10 containing the binding energy upon using COSMO was investigated (see Figure 4 and Table S6 in SI). From this it is obvious that the most correlated and contributing descriptor to the binding energy is the anisotropy ($P_3$). This indicates an impact of the SAA-SOM-interaction on the orientation of the interacted SOM fragment to SAA. The other descriptors that include the dipole moment ($P_1$), quadrupole moment ($P_2$), sum of the partial charges on O atoms ($P_4$), sum of the partial charges on C+O+N atoms ($P_5$), and molecular-mass ($P_6$) are significantly correlated to the binding energy as well.

To the best of our knowledge this is the first study revealing a quantum mechanical molecular level picture for the interaction of complex, multifunctional SOM with polar and non-polar organic pollutants. Particularly novel is the demonstrated agreement between the experimental (mass spectrometric SOM characterization and sorption experiments) and theoretical (calculated binding energies and estimated Freundlich unit capacities in gas and solution phase, QSAR) outcomes; thus this approach is recommended as an efficient tool for studying interactions of a wide range of organic pollutants with SOM.

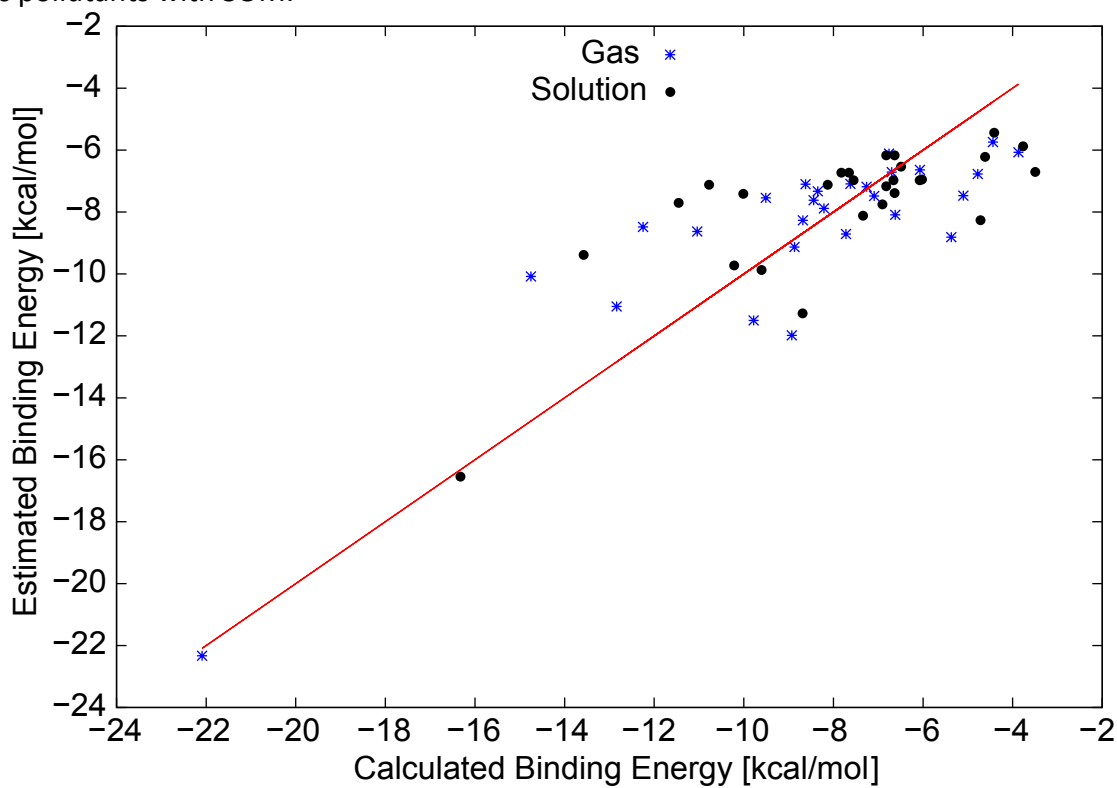

Figure 4. The estimated binding energies of SAA with the SOM systems versus the calculated ones at B3LYP/D3/6-311++G(d,p) in gas phase (blue) as well as in solution (black). The red line is a linear correlation (r = 1.00) that was plotted as guide for the eye.

## 4.   Conclusions

Understanding of soil-related processes at the molecular level remains to be a challenge of environmental science. Many of the complications arise from the heterogeneous nature of SOM, which makes it difficult to relate the molecular to the macroscopic level. With the present work this issue is tackled using a previously developed molecular test set containing functionalities representative for SOM. Thereby, we extended our approach towards sulfonamides, an important class of pharmaceutical antibiotics and highly polar chemicals. Efficient equations correlating the experiment with the theory were established providing the ability to estimate the Freundlich unit capacity from the calculated binding energy. This evidences that our previously developed SOM model is flexible enough to describe the SAA-SOM-interactions.

For the case of SAA the following results have been obtained. Experimentally, it was shown that sorption of SAA to soil is more closely correlated to the SOM content as compared to the content and composition of soil minerals which are largely divers across particle-size fractions. Moreover, it was shown that the SAA-SOM-interaction depends on the chemical composition of SOM more than the SOM content. Although SAA binds to both hydrophilic and hydrophobic interaction sites, both experiment and theory showed that SAA obeys a site-specific sorption on the soil surfaces. Due to the interaction of SAA with the SOM molecular systems, SAA formed H-bonds with the polar interaction sites while the dispersion interaction dominates for the non-polar interaction sites. More specifically, for soils of low water content (simulated via the gas phase calculations), SAA binds to the SOM molecular systems in the order cationic species > anionic species > peptides > carbohydrates > phenols and lignin monomers > lignin dimers > N-containing heterocyclic compounds > fatty acids > sterols > alkylated aromatic compounds > lipids, alkanes, and alkenes. By increasing the amount of water supplied to soils (simulated via the COSMO calculations), the binding strength for SAA to the SOM molecular systems decreased and the order of binding changed except for peptides and charged systems. Differences in the binding energies due to changes in the molecular structure were found. Compared to the hydrophobic compound hexachlorobenzene (HCB) it can be stated that the adsorption of the hydrophilic SAA is weaker. Subsequently, the SAA-SOM binding has been scrutinized using the QSAR approach. It highlighted the significant role of the polarity, partial charges on the electronegative atoms, orientation, and molecular volume of the interacting SOM fragment with SAA in the SAA-SOM binding process. Finally, having successfully investigated polar as well as non-polar organic pollutants and validated the experimental-theoretical approach by a good agreement between the total binding energies, calculated from the partial binding energies of representative test set molecules and their quantity in Py-FI mass spectra, and the experimentally determined Freundlich unit capacities, we can conclude that our SOM model can be recommend as a tool for studying interactions of a wide range of organic pollutants with SOM.

## Acknowledgment

Ashour A. Ahmed acknowledges STZ Soil Biotechnology (Steinbeis GmbH & Co. KG für Technologietransfer), Stuttgart, Germany. Financial support for the sorption experiments was provided by the Deutsche Forschungsgemeinschaft (DFG, Th 678/3-1), Bonn, Germany. The assistance of M.-O. Aust in conducting the soil fractionation and sorption experiments is gratefully acknowledged.

## Supplementary information

This part contains more details about the previous approaches for SOM modeling, particle-size fractionation, Py-FIMS, QSAR (method description as well as results and discussion), and finally, the effect of the dispersion and the soil solution on the SAA-SOM-interaction. This supplementary information involves 7 Figures and 6 Tables.